\documentclass[a4paper]{article}

\usepackage[american]{babel}
\usepackage[dvips]{graphicx}
\usepackage{amsmath,amsfonts,amsthm,mathrsfs}
\usepackage{a4wide}

\newcommand{\abs}[1]{\vert #1\vert}
\newcommand{\asy}{\text{asympt}}
\newcommand{\B}{\mathcal{B}}
\newcommand{\cE}{\mathcal{E}}
\newcommand{\dbs}{d_\textup{BS}}
\newcommand{\dsum}{\displaystyle\sum}
\newcommand{\E}{\mathbb{E}}
\newcommand{\floor}[1]{\left\lfloor #1\right\rfloor}
\newcommand{\intl}{\int\limits}
\newcommand{\iintl}{\iint\limits}
\newcommand{\norm}[1]{\Vert #1\Vert}
\newcommand{\R}{\mathbb{R}}
\newcommand{\Tmax}{T_\textup{max}}
\newcommand{\uu}{\mathbf{u}}

\theoremstyle{remark}\newtheorem{remark}{Remark}

\graphicspath{{./}{figure/}}

\title{On the dynamics of social conflicts: \\ looking for the Black Swan}
\author{Nicola Bellomo\textsuperscript{a},
        Miguel A. Herrero\textsuperscript{b},
        Andrea Tosin\textsuperscript{c} \\[0.5cm]
    {\small\it\textsuperscript{a}Department of Mathematics, Politecnico di Torino}\\[-1mm]
    {\small\it Corso Duca degli Abruzzi 24, 10129 Torino, Italy}\\[-1mm]
    {\small\it nicola.bellomo@polito.it}\\[1mm]
    {\small\it\textsuperscript{b}Department of Applied Mathematics, Universidad Complutense}\\[-1mm]
    {\small\it Plaza de Ciencias 3, Ciudad Universitaria, 28040 Madrid, Spain}\\[-1mm]
    {\small\it Miguel\_Herrero@mat.ucm.es}\\[1mm]
    {\small\it\textsuperscript{c}Istituto per le Applicazioni del Calcolo ``M. Picone''}\\[-1mm]
    {\small\it Consiglio Nazionale delle Ricerche}\\[-1mm]
    {\small\it Via dei Taurini 19, 00185 Roma, Italy}\\[-1mm]
    {\small\it a.tosin@iac.cnr.it}
}
\date{}

\begin{document}
\maketitle

\begin{abstract}
This paper deals with the modeling of social competition, possibly resulting in the onset of  extreme conflicts. More precisely, we discuss models describing the interplay between individual competition for wealth distribution that, when coupled with political stances coming from support or opposition to a government, may give rise to strongly self-enhanced effects. The latter may be thought of as the early stages of massive, unpredictable events known as Black Swans, although no analysis of any fully-developed Black Swan is provided here. Our approach makes use of the framework of the kinetic theory for active particles, where nonlinear interactions among subjects are modeled according to game-theoretical tools.

\medskip
\end{abstract}

\section{Introduction}

The dynamics of social and economic systems are necessarily based on individual behaviors, by which single subjects express, either consciously or unconsciously, a particular strategy, which is heterogeneously distributed. The latter is often based not only on their own individual purposes, but also on those they attribute to other agents. However, the sheer complexity of such systems makes it often difficult to ascertain the impact of personal decisions on  the resulting collective dynamics. In particular, interactions among individuals need not have an additive, linear character. As a consequence, the global impact of a given number of entities (``field entities'') over a single one (``test entity'') cannot be assumed to merely consist in  the linear superposition of any single field entity action. This nonlinear feature represents a serious conceptual difficulty to the derivation, and subsequent analysis, of mathematical models for that type of systems.

In the last few years, a radical philosophical change has been undertaken in social and economic disciplines. An interplay among Economics, Psychology, and Sociology has taken place, thanks to a new cognitive approach no longer grounded on the traditional assumption of rational socio-economic behavior. Starting from the concept of bounded rationality \cite{simon1959tdm}, the idea of Economics as a subject highly affected by individual (rational or irrational) behaviors, reactions, and interactions has begun to impose itself. In this frame, the contribution of mathematical methods to a deeper understanding of the relationships between individual behaviors and collective outcomes may be fundamental. All of these concepts are expressed in the PhD dissertation \cite{ajmone2009npm}, that the interested reader is referred to also for additional pertinent bibliography. 

More in general in fields ranging from Economics to Sociology and Ecology, the last decades have witnessed an increasing interest for the introduction of quantitative mathematical methods that could  account for individual, not necessarily rational, behaviors. Terms as game theory, bounded rationality, evolutionary dynamics are often used in that context and clearly illustrate the  continuous search  for techniques able to provide mathematical models than can describe, and predict,  living behaviors, see \cite{camerer2003bgt,camerer2003abe,simon1982mbr-2,simon1982mbr-1,simon1997mbr-3}, as also documented in the bibliography on evolutionary game theory  cited in the following. As a result, a picture of social and biological sciences as evolutionary complex systems is unfolding \cite{arlotti2012cid,bellouquid2012mvt}. A key experimental feature of such systems is that interaction among heterogeneous individuals often produces unexpected outcomes, which were absent at the individual level, and are commonly termed emergent behaviors. 

The new point of view promoted the image of Economics as an evolving complex system, where interactions among heterogeneous individuals produce unpredictable emerging outcomes \cite{arthur1997eec,kirman2000lbl}. In this context, setting up a mathematical description able to capture the evolving features of socio-economic systems is a challenging, however difficult, task, which calls for a proper interaction between mathematics and social sciences. In this paper, a preliminary step in this direction is attempted. A mathematical framework is outlined, suitable to incorporate some of the main complexity features of socio-economic systems. Out of it, specific mathematical models are derived, focusing in particular on the prediction of the so called \emph{Black Swan}. The latter is defined to be a rare event, showing up as an irrational collective trend generated by possibly rational individual behaviors \cite{taleb2007bsi,taleb2010ffr}.

To achieve our goal, we will use mathematical tools based on a development of the kinetic theory for active particles, see e.g., \cite{arlotti2002gkb,bellomo2010mhl,bellomo2011mtl}, suitable to include nonlinear interactions and learning phenomena. The hallmarks of the approach can be summarized as follows: the system is partitioned into \emph{functional subsystems}, whose entities, called \emph{active particles}, are characterized by an individual state termed \emph{activity}; the state of each functional subsystem is defined by a probability distribution over the activity variable; interactions among active particles, generally nonlocal and nonlinearly additive, are treated as \emph{stochastic games}, meaning that the pre-interaction states of the particles and the post-interaction ones can be known only in probability; finally, the evolution of the probability distribution is obtained by a balance of particles within elementary volumes of the space of microscopic states, the inflow and outflow of particles being related to the aforementioned interactions. A general theory for linearly additive interactions, along with various applications, is reported in \cite{bellomo2008mcl}, whereas a first extension to non-additive and nonlocal interactions, modeled by methods of the stochastic game theory, is included in \cite{bellomo2012mts}.

This mathematical approach has been applied to various fields of Life Sciences, such as social systems
\cite{bertotti2008cla}, opinion formation \cite{bertotti2008dgk}, and has been revisited in
\cite{ajmone2009msc,ajmone2008mtc} with reference to Behavioral Sciences including Politics and Economics. Moreover, it
has also been applied in fields different from the above-mentioned ones, for instance propagation of epidemics under
virus mutations \cite{delillo2009mev,delitala2011mmk} and theory of evolution \cite{bellomo2011mtl}. In all of these
applications, the heterogeneous behavior of individuals and random mutations are important features characterizing the
systems under consideration. The conceptual link between methods of statistical mechanics and game theory was also introduced by Helbing \cite{helbing2010qsd}; on the other hand, also methods of the mean-field kinetic theory have been used to model socio-economic systems, see e.g., \cite{during2009bfp,toscani2006kmo}. The specialized literature offers a great variety of different approaches, such as population dynamics with structure \cite{webb1985tna} or super-macroscopic dynamical systems \cite{nuno2011mmc}. In all cases, the challenging goal to be met consists in capturing the relevant features of living complex systems.

After the above general overview, the plan of the paper can now be illustrated in more detail. The contents are distributed into four more sections. Section~\ref{sect:compl.asp} analyzes the complexity aspects of socio-economic systems, in particular five key features are selected to be retained in the modeling approach. Section~\ref{sect:compl.red} introduces the mathematical structures of the kinetic theory for active particles, which offer the basis for the derivation of specific models. Section~\ref{sect:case.studies} opens with two illustrative applications focused on social conflicts: the first one shows that a social competition, if not properly controlled, may induce an unbalanced distribution of wealth with a clustering of the population in two extreme classes (a large class of poor people and a small oligarchic class of wealthy ones); the second one exemplifies, in connection with the aforesaid dynamics, how such a clustering can lead to a growing opposition against a government. Subsequently, the investigation moves on the identification of premonitory signals, that can provide preliminary insights into the emergence of a Black Swan viewed as a large deviation from some heuristically expected trend. Section~\ref{sect:discussion} finally proposes a critical analysis and focuses on research perspectives.

\section{Complexity aspects of socio-economic systems}
\label{sect:compl.asp}
In this section the complexity features of socio-economic systems are analyzed, with the aim of extracting some
hallmarks to be included in mathematical models.

Socio-economic systems can be described as ensembles of several living entities, viz. active particles, whose individual behaviors simultaneously affect and are affected by the behaviors of a certain number of other particles. These actions depend, in most cases, on the number of interacting particles, their localization, and their state. Generally, individual actions are rational, focused on a well-defined goal, and aimed at individual benefit. On the other hand, some particular situations may give rise to behaviors in contrast with that primary goal, like e.g., in case of panic. A further aspect to be considered is that the system is generally composed by parts, which are interconnected and interdependent. Namely, every system is formed by nested subsystems, so that interactions occur both within and among subsystems.

As a matter of fact, all living systems exhibit some common features, whereas others can vary depending on the type of system under consideration. We will assume that all subsystems of a given system are characterized by the same features, possibly expressed with larger or smaller intensity according to their specificity.

Bearing all above in mind, in the following some specific aspects of socio-economic systems, understood as living complex systems, are identified and commented. The selection is limited to five features, in order to avoid an over-proliferation of concepts, considering that mathematical equations cannot include the whole variety of complexity issues. Thus, the list below does not claim to be exhaustive, rather it is generated by the authors' personal experience and bias.
\begin{enumerate}
\item \textbf{Emerging collective behaviors}. Starting from basic individual choices, interaction dynamics produce the spontaneous emergence of collective behaviors, that, in most cases, are completely different or apparently not contained in those of the single active particles. Ultimately, \emph{the whole can be much more than the sum of its parts}. This is possible because active particles typically operate out of equilibrium.
\item \textbf{Strategy, heterogeneity, and stochastic games}. Active particles have the ability to develop specific strategies, which depend also on those expressed by the other particles. Normally, such strategies are generated by rational principles but are heterogeneously distributed among the particles. Furthermore, irrational behaviors cannot be excluded. Accordingly, the representation of the system needs random variables, and interactions have to be modeled in terms of stochastic games because it is not possible to identify an average homogeneous rational attitude.
\item \textbf{Nonlinear and nonlocal interactions}. Interactions among active particles are generally nonlinear and nonlocal, because they depend on the global distribution of some close and/or far neighbors. The latter have to be identified in terms of a suitable distance among the microscopic states of the particles. Active particles play a game at each interaction: the outcome, which depends nonlinearly on the states of all interacting particles, modifies their state in a stochastic manner.
\item \textbf{Learning and evolution}. Individuals in socio-economic systems are able to learn from their experience. This implies that the expression of the strategy evolves in time, and consequently that interaction dynamics undergo modifications. In some cases, special situations (e.g., onset of panic) can even induce quick modifications. Moreover, adaptation to environmental conditions and search of one's own benefit may induce mutations and evolution.
\item \textbf{Large number of components}. Living systems are often constituted by a great deal of individual diversity, such that a detailed description focused on single active particles would be actually infeasible. Therefore, a complexity reduction, by means of suitable mathematical strategies, is necessary for handling them at a practical level.
\end{enumerate}

\section{Complexity reduction and mathematical tools}
\label{sect:compl.red}
This section provides the conceptual lines leading to the methods of the kinetic theory for active particles, which has been selected as the mathematical framework suitable to derive specific models here. The presentation is followed by a critical analysis aimed at checking the consistency of the mathematical approach with the issues discussed in Section~\ref{sect:compl.asp}, as well as its efficiency in reducing the complexity of the real system. Modeling is concerned with systems of interacting individuals belonging to different groups. Their number is supposed to be constant in time, namely birth and death processes or inlets from an outer environment are not taken into account.

It is worth stressing that the approach used in the various papers cited in the Introduction was based on linearly additive interactions with parameters constant in time. Here, on the contrary, both nonlinearly additive interactions and time-evolving parameters, due to the conditioning by the collective state of the system, are considered. We recall that interactions are said to be \emph{linearly additive} when the outcome of each of them is not influenced by the presence of particles other than the interacting ones, so that a superposition principle holds true: the action on a particle is the sum of all actions applied individually by the other particles. For instance, this is the case of mean field theories. Otherwise, interactions are said to be \emph{nonlinearly additive}. We point out that the progress from linear to nonlinear interactions is crucial in the attempt of inserting into mathematical equations individual behaviors that may change quickly, possibly under the influence of the outer environment. Nonlinearities can be generated in several ways. In particular, the particles which play a role in the interactions may be selected by means of rational actions, whereas the interaction output can be obtained from individual strategies and interpretation of actions exerted vby other particles. These concepts are well understood in the interpretation of swarming phenomena from the point of view of both Physics \cite{ballerini2008ira} and Mathematics \cite{bellomo2012mts,bellouquid2012mvt,cristiani2011eai}.

\subsection{Active particles, heterogeneity, functional subsystems, and representation issues}
As already stated, the living entities of the system at hand will be regarded as particles able to express actively a certain social and/or economic strategy based on their socio-economic state. Such a strategy will be called \emph{activity}. In general, the system might be constituted by different types of active particles, each of them featuring a different strategy. However, aiming at a (necessary) complexity reduction, the system can be decomposed in \emph{functional subsystems} constituted by active particles that individually express the same strategy. In other words, whenever the strategy of the active particles of a system is heterogeneous, the modeling approach should identify an appropriate decomposition into functional subsystems, within each of which the strategy is instead homogeneous across the member active particles.

In each functional subsystem, the microscopic activity, denoted by $u$, can be taken as a scalar variable belonging to a domain $D_u\subseteq\R$. In some cases, it may be convenient to assume that $D_u$ coincides with the whole $\R$.

Let us consider a decomposition of the original system into $m\geq 1$ functional subsystems labeled by an index $p=1,\,\dots,\,m$. According to the kinetic theory for active particles, each of them is described by a time-evolving distribution function over the microscopic activity $u$:
$$ f^p=f^p(t,\,u):[0,\,\Tmax]\times D_u\to\R_+, $$
$\Tmax>0$ being a certain final time (possibly $+\infty$), such that the quantity $f^p(t,\,u)\,du$ is the (infinitesimal) number of active particles of the $p$-th subsystem having at time $t$ an activity comprised in the (infinitesimal) interval $[u,\,u+du]$.

Under suitable integrability conditions, the number of active particles in the $p$-th functional subsystem at time $t$ is
$$ N^p(t):=\intl_{D_u}f^p(t,\,u)\,du. $$
More in general, it is possible to define (weighted) moments of any order $l$ of the distribution functions\footnote{Notice that, in this formula, $l$ is a true exponent whereas $p$ is a superscript.}:
$$ \E^p_l(t):=\intl_{D_u}u^lf^p(t,\,u)w(u)\,du, $$
where $w:D_u\to\R_+$ is an appropriate weight function with unit integral on $D_u$. $\E^p_l(t)$ can be either finite or infinite, according to the integrability properties of $f^p(t,\,\cdot)$.

If the number of particles within each subsystem is constant in time, so that no particle transition occurs among subsystems, then each $f^p$ can be normalized with respect to $N^p(0)$ and understood as a probability density. Alternatively, it is possible to normalize with respect to the total number of particles of the system $\sum_{p=1}^{m}N^p(0)$, which entails
$$ \sum_{p=1}^{m}\intl_{D_u}f^p(t,\,u)\,du=1, \quad \forall\,t\in[0,\,\Tmax]. $$

It has been shown in \cite{bertotti2008dgk}, see also the references cited therein, that for various applications it is convenient to assume that the activity is a discrete variable, especially when \emph{activity classes} can be more readily identified in the real system. A lattice $I_u=\{u_1,\,\dots,\,u_i,\,\dots,\,u_n\}$ is thus introduced in the domain $D_u$, admitting that $u\in I_u$. The representation of the $p$-th functional subsystem is now provided by a set of $n\geq 1$ distribution functions
$$ f^p_i=f^p_i(t):[0,\,\Tmax]\to\R_+, \quad i=1,\,\dots,\,n $$
such that $f^p_i(t)$ is the (possibly normalized) number of active particles in the $i$-th activity class of the $p$-th subsystem at time $t$. Formally, we have $f^p_i(t)=f^p(t,\,u_i)$ or, in distributional sense,
$$ f^p(t,\,u)=\sum_{i=1}^{n}f^p_i(t)\delta_{u_i}(u), $$
where $\delta_{u_i}$ is the Dirac distribution centered at $u_i$  . The formulas given above for the moments of the distribution remain valid, provided integrals on $D_u$ are correctly understood as discrete sums over $i$:
\begin{equation}
    N^p(t)=\sum_{i=1}^{n}f^p_i(t), \qquad \E^p_l(t)=\sum_{i=1}^{n}u_i^lf^p_i(t)w(u_i).
    \label{eq:moments.discr}
\end{equation}

\subsection{Interactions, stochastic games, and collective dynamics}
\label{sect:int.stochgam.colldyn}
In general, interactions involve particles of both the same and different functional subsystems. In some cases, interactions between active particles and the outer environment have also to be taken into account. The outer environment is typically assumed to have a known state, which is not modified by interactions with the system at hand. In particular, in this paper we consider the simple case of systems which do not interact with the outer environment, besides possibly an action applied by the latter that modifies some interaction rules.

The description of the interactions can essentially be of the following two types: \emph{deterministic} if the output is univocally identified given the states of the interacting entities (generally related to  standard rational behavior of the particles, when large deviations are not expected); \emph{stochastic} if the output can be known only in probability, due for instance to a variability in the reactions of the particles to similar conditions. In the latter case, interactions are understood as \emph{stochastic games}. In the present context, our interest is mainly in stochastic games because of possible irrational behaviors, as outlined in Section~\ref{sect:compl.asp}.

When describing interactions among active particles, it is useful to distinguish three main actors named test, candidate, and field particles. This terminology is indeed standard in the kinetic theory for active particles.
\begin{itemize}
\item The \textbf{test} particle, with activity $u$, is a generic representative entity of the functional subsystem under consideration. Studying interactions within and among subsystems means studying how the test particle can loose its state or other particles can gain it.
\item \textbf{Candidate} particles, with activity $u_\ast$, are the particles which can gain the test state $u$ in consequence of the interactions.
\item \textbf{Field} particles, with activity $u^\ast$, are the particles whose presence triggers the interactions of the candidate particles.
\end{itemize}

The modeling of the interactions is based on the derivation of two terms:  the \emph{interaction rate} and the
\emph{transition probabilities}. Let us consider, separately, some preliminary guidelines for their construction.

\paragraph*{Interaction rate} This term, denoted by $\eta^{pq}$, models the frequency of the interactions between candidate and field particles belonging to the $p$-th and $q$-th functional subsystems, respectively.

\paragraph*{Transition probabilities} The general rule to be followed in modeling stochastic games is that candidate particles can acquire, in probability, the state of the test particle after an interaction with field particles, while the test particle can lose, in probability, its own. Such dynamics are described by the transition probabilities $\B^{pq}$, which express the probability that a candidate particle of the $p$-th subsystem ends up into the state of the test particle (of the same subsystem) after interacting with a field particle of the $q$-th subsystem.

In case of linear interactions, $\B^{pq}$ is conditioned only by the states of the interacting particles for each pair of functional subsystems: $\B^{pq}=\B^{pq}(u_\ast\to u\vert u_\ast,\,u^\ast)$. In addition, it satisfies the following condition:
\begin{equation}
    \intl_{D_u}\B^{pq}(u_\ast\to u\vert u_\ast,\,u^\ast)\,du=1, \quad \forall\,u_\ast,\,u^\ast\in D_u, \quad
        \forall\,p,\,q=1,\,\dots,\,m,
    \label{eq:sum.prob}
\end{equation}
which, in case of discrete activity, becomes
\begin{equation}
    \sum_{i=1}^{n}\B^{pq}_{hk}(i)=1, \quad \forall\,h,\,k=1,\,\dots,\,n, \quad
        \forall\,p,\,q=1,\,\dots,\,m,
    \label{eq:sum.prob.discr}
\end{equation}
where we have denoted $\B^{pq}_{hk}(i):=\B^{pq}(u_h\to u_i\vert u_h,\,u_k)$.

Nonlinear interactions imply, instead, that particles are not simply subject to  the superposition of binary actions but are also affected by the global current state of the system. Consequently, $\B^{pq}$ may be conditioned by the moments of the distribution functions. Denoting by $\cE^p_L=\{\E^p_l\}_{l=1}^{L}$ the set of all moments of the distribution function $f^p$ up to some order $L\geq 0$, the formal expression of the transition probabilities is now $\B^{pq}=\B^{pq}(u_\ast\to u\vert u_\ast,\,u^\ast;\,\cE^p_L,\,\cE^q_L)$, along with a condition analogous to that expressed by Eq.~\eqref{eq:sum.prob}. In case of discrete activity, rather than introducing a new notation we simply redefine $\B^{pq}_{hk}(i):=\B^{pq}(u_h\to u_i\vert u_h,\,u_k;\,\cE^p_L,\,\cE^q_L)$, so as to avoid an over-proliferation of symbols.

\medskip

The above models of interactions lead straightforwardly to the derivation of a system of evolution equations for the set of distribution functions $\{f^p\}_{p=1}^m$, obtained from a balance of incoming and outgoing fluxes in the elementary volume $[u,\,u+du]$ of the space of microscopic states. The resulting mathematical structure, to be used as a paradigm for the derivation of specific models, is as follows:
\begin{align}
   & \frac{\partial f^p}{\partial t}(t,\,u)=
     \sum_{q=1}^{m}\iintl_{D_u^2}\eta^{pq}(t,\,u_\ast,\,u^\ast)
        \B^{pq}(u_\ast\to u\vert u_\ast,\,u^\ast;\,\cE^p_L,\,\cE^q_L)
                f^p(t,\,u_\ast)f^q(t,\,u^\ast)\,du_\ast\,du^\ast \nonumber \\
    & \qquad - f^p(t,\,u)\sum_{q=1}^{m}\intl_{D_u}\eta^{pq}(t,\,u,\,u^\ast)f^q(t,\,u^\ast)\,du^\ast,
        \qquad p=1,\,\dots,\,m,
    \label{eq:evol.cont}
\end{align}
which, in case of discrete activity, formally modifies as
\begin{equation}
    \frac{df^p_i}{dt}(t) = \sum_{q=1}^{m}\sum_{k=1}^{n}\sum_{h=1}^{n}\eta^{pq}_{hk}(t)
        \B^{pq}_{hk}(i)f^p_h(t)f^q_k(t)-f^p_i(t)\sum_{q=1}^{m}\sum_{k=1}^{n}\eta^{pq}_{ik}(t)f^q_k(t)
            \label{eq:evol.disc}
\end{equation}
for $i=1,\,\dots,\,n$ and $p=1,\,\dots,\,m$.

\subsection{Interactions with transitions across functional subsystems}
\label{sect:transitions}
The mathematical framework presented in the preceding section does not account for changes of functional subsystem by the active particles. However, transitions across subsystems may be relevant in modeling concomitant social and economic dynamics, particularly if the various subsystems can be related to different aspects of the microscopic state of the active particles.

To be more specific, and to anticipate the application that we will be concerned with in the next sections, consider the case of a vector activity variable $\uu=(u,\,v)\in D_u\times D_v\subseteq\R^2$, with $u$ representing the economic state of the active particles of a certain country and $v$ their level of support/opposition to the government policy. In order to reduce the complexity of the system, we will assume that the component $v$ of the microscopic state is discrete: $v\in I_v=\{v_1,\,\dots,\,v_p,\,\dots,\,v_m\}\subset D_v$, and we will use the lattice $I_v$ as a criterion for identifying the functional subsystems. In practice, each subsystem gathers individuals expressing a common opinion on the government's doings. It is plain that, in order to obtain an accurate picture of the interconnected socio-economic dynamics, besides economic interactions within and among subsystems, transitions of active particles across the latter have also to be considered.

To this end, the mathematical structures previously derived need to be duly generalized. Specifically, the transition probabilities read now
$$ \B^{pq}(r)=\B^{pq}(r)(u_\ast\to u\vert u_\ast,\,u^\ast;\,\cE^p_L,\,\cE^q_L,\,\cE^r_L), \quad
    u,\,u_\ast,\,u^\ast\in D_u, \quad p,\,q,\,r=1,\,\dots,\,m $$
for expressing the probability that a candidate particle of the $p$-th subsystem with activity $u_\ast$ ends up into the $r$-th subsystem with activity $u$ after an interaction with a field particle of the $q$-th subsystem with activity $u^\ast$. Notice that, in case of nonlinearly additive interactions, these probabilities are generally conditioned also by the moments of the distribution functions of the output subsystem. The new transition probabilities satisfy the normalization condition:
$$ \sum_{r=1}^{m}\intl_{D_u}\B^{pq}(r)(u_\ast\to u\vert u_\ast,\,u^\ast;\,\cE^p_L,\,\cE^q_L,\,\cE^r_L)\,du=1,
        \quad \forall\,u_\ast,\,u^\ast\in D_u,\,p,\,q=1,\,\dots,\,m. $$

Following the same guidelines that led to the derivation of the mathematical structures of Section~\ref{sect:int.stochgam.colldyn}, we obtain the new equations with transitions across subsystems as
\begin{align}
    \frac{\partial f^r}{\partial t}(t,\,u)&=\sum_{p=1}^{m}\sum_{q=1}^{m}\iintl_{D_u^2}\eta^{pq}(t,\,u_\ast,\,u^\ast)
        \B^{pq}(r)(u_\ast\to u\vert u_\ast,\,u^\ast;\,\cE^p_L,\,\cE^q_L,\,\cE^r_L) \nonumber \\
    & \phantom{=} \qquad\qquad\quad\times f^p(t,\,u_\ast)f^q(t,\,u^\ast)\,du_\ast\,du^\ast \nonumber \\
    & \phantom{=} -f^r(t,\,u)\sum_{q=1}^{m}\intl_{D_u}\eta^{rq}(t,\,u,\,u^\ast)f^q(t,\,u^\ast)\,du^\ast,
        \qquad r=1,\,\dots,\,m.
    \label{eq:evol.cont.2}
\end{align}
It is worth noticing that by putting
$$ \B^{pq}(r)(u_\ast\to u\vert u_\ast,\,u^\ast;\,\cE^p_L,\,\cE^q_L,\,\cE^r_L)=
    \B^{pq}(u_\ast\to u\vert u_\ast,\,u^\ast;\,\cE^p_L,\,\cE^q_L)\delta_{pr}, $$
where $\delta_{pr}=1$ if $p=r$, $\delta_{pr}=0$ otherwise, one recovers from Eq.~\eqref{eq:evol.cont.2} the particular case of interactions without transitions across subsystems described by Eq.~\eqref{eq:evol.cont}.

Models relying on Eq.~\eqref{eq:evol.cont.2} are \emph{hybrid}, because the economic state $u$ is treated as a continuous variable whereas the decomposition in functional subsystem, linked to socio-political beliefs, is of a discrete nature. Correspondingly, the space of microscopic states is $D_u\times I_v$. If also the variable $u$ is discrete within each subsystem then the space of microscopic states is the full lattice $I_u\times I_v$ and Eq.~\eqref{eq:evol.cont.2} reads
\begin{align}
    \frac{df^r_i}{dt}(t) &= \sum_{p=1}^{m}\sum_{q=1}^{m}\sum_{k=1}^{n}\sum_{h=1}^{n}\eta^{pq}_{hk}(t)
        \B^{pq}_{hk}(i,\,r)f^p_h(t)f^q_k(t) \nonumber \\
    & \phantom{=} -f^r_i(t)\sum_{q=1}^{m}\sum_{k=1}^{n}\eta^{rq}_{ik}(t)f^q_k(t), \quad
        i=1,\,\dots,\,n, \quad r=1,\,\dots,\,m,
    \label{eq:evol.disc.2}
\end{align}
where $\B^{pq}_{hk}(i,\,r):=\B^{pq}(r)(u_h\to u_i\vert u_h,\,u_k;\,\cE^p_L,\,\cE^q_L,\,\cE^r_L)$ fulfills
$$ \sum_{r=1}^{m}\sum_{i=1}^{n}\B^{pq}_{hk}(i,\,r)=1, \quad \forall\,h,\,k=1,\,\dots,\,n, \quad
    \forall\,p,\,q=1,\,\dots,\,m. $$
Equation~\eqref{eq:evol.disc.2} is a generalization of Eq.~\eqref{eq:evol.disc}, which is recovered as a particular case by letting $\B^{pq}_{hk}(i,\,r)=\B^{pq}_{hk}(i)\delta_{pr}$.

\section{On the interplay between socio-economic dynamics and political conflicts}
\label{sect:case.studies}
The theory developed in the previous sections provides a background for tackling some illustrative applications concerned with social competition problems. Particularly, our main interest here lies in phenomena such as unbalanced wealth distributions possibly leading to popular rebellion against governments, which can be classified as Black Swans. The envisaged scenario shares some analogies with the events recently observed in North Africa countries, though in a simplified context. The mathematical models, indeed, are going to be some minimal exploratory ones, with a small number of functional subsystems and parameters for describing interactions at the microscopic scale.

We focus on closed systems, such as a country with no interactions with other countries featuring similar political and/or religious organizations. A natural goal to pursue is then understanding which kind of interactions among different socio-economic classes of the same country can produce the aforesaid Black Swans. It is worth mentioning that, when interactions with other countries are considered, the investigation can also address propagation by a domino effect.

The contents of the following subsections organize the previous ideas through three steps. The first one refers to welfare dynamics in terms of cooperation and competition among economic classes. The second one focuses instead on dynamics of support and opposition to a certain regime triggered by the welfare distribution. Finally, the third one proposes a preliminary approach to the identification of premonitory signals possibly implying the onset of a Black Swan, here understood as an exceptional growth of opposition to the regime fostered by the synergy with socio-economics dynamics.

In order to model the above-mentioned cooperation/competition interactions, we adopt the following qualitative paradigm of consensus/dissensus dynamics:
\begin{itemize}
\item \emph{Consensus} -- The candidate particle sees its state either increased, by profiting from a field particle with a higher state, or decreased, by pandering to a field particle with a lower state. After mutual interaction, the states of the particles become closer than before the interaction.
\item \emph{Dissensus} -- The candidate particle sees its state either further decreased, by facing a field particle with a higher state, or further increased, by facing a field particle with a lower state. After mutual interaction, the states of the particles become farther than before the interaction.
\end{itemize}
Once formalized at a quantitative level, this paradigm can act as a base for constructing the transition probabilities introduced in Section~\ref{sect:compl.red}.

\begin{remark}
Although the modeling of the transition probabilities relates to the microscopic interactions among active particles, it is worth mentioning that recent contributions to game theory, especially the approach to evolutionary games presented in \cite{gintis2009gte,helbing2010qsd,nowak2006ede,nowak2004edb,santos2006eds,santos2012edc}, address interactions at the macroscopic scale. In such a context, learning abilities and evolution are essential features of the modeling strategy, while changes in the external environment can induce modifications of rational behaviors up to irrational ones.
\end{remark}

\subsection{Modeling socio-economic competition}
\label{sect:welfare}
In this section we consider the modeling of socio-economic interactions based on the  previously discussed consensus/dissensus dynamics. This problem has been first addressed in \cite{bertotti2008cla} for a large community of individuals divided into different social classes. The model proposed by those authors introduces a critical distance, which triggers either cooperation or competition among the classes. In more detail, if the actual distance between the interacting classes is lower than the critical one then a competition takes place, which causes a further enrichment of the wealthier class and a further impoverishment of the poorer one. Conversely, if the actual distance is greater than the critical one then the social organization forces cooperation, namely the richer class has to contribute to the wealth of the poorer one.

\begin{figure}[t]
\centering
\includegraphics[width=0.7\textwidth]{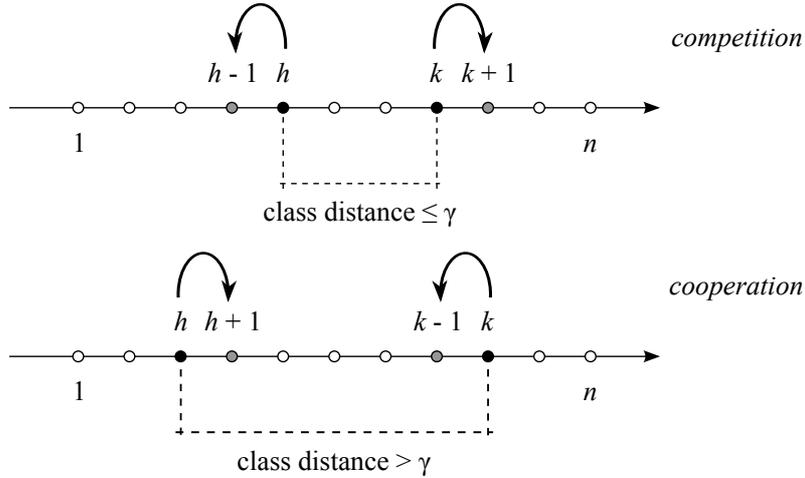}
\caption{Dynamics of competition (top) and cooperation (bottom) between pairs of candidate (index $h$) and field (index $k$) active particles. The critical distance $\gamma$, which triggers either behavior depending on the actual distance between the interacting classes, may evolve in time according to the global evolution of the system.}
\label{fig:coop-comp_dyn}
\end{figure}

In the above-cited paper, linearly additive interactions are used along with a constant critical distance. Such an approach is here revisited by introducing nonlinearly additive interactions and a critical distance which evolves in time depending on the global wealth distribution. In more detail, the characteristics of the present framework are summarized as follows.
\begin{itemize}
\item \emph{Functional subsystems}. A single functional subsystem ($m=1$) is considered, constituted by the population of a country or of a regional area. For the sake of convenience, in this case we drop any superscript referring to functional subsystems ($p,\,q,\,\dots$).
\item \emph{Activity}. The activity variable $u$ identifies the wealth status of the active particles.
\item \emph{Encounter rate}. Two different rates of interactions are considered, corresponding to competitive and cooperative interactions, respectively.
\item \emph{Strategy leading to the transition probabilities}. When interacting with other particles, each active particle plays a game with stochastic output. If the difference of wealth class between the interacting particles is lower than a critical distance $\gamma$ then the particles compete in such a way that those with higher wealth increase their state against those with lower wealth. Conversely, if the difference of wealth class is higher than $\gamma$ then the opposite occurs (see Fig.~\ref{fig:coop-comp_dyn}). The critical distance evolves in time according to the global wealth distribution over wealthy and poor particles. It may be influenced, at least partially, by the social policy of the government, to be regarded in the present application as an external action.
\end{itemize}

This modeling approach can be developed for both continuous and discrete activity variable. The specific model proposed here is derived assuming that the activity is a discrete variable, which, as observed in \cite{bertotti2008elc}, allows one to identify the microscopic states of the population by ranges, namely by a finite number $n$ of classes $u_1,\,\dots,\,u_n$. This is not only practical from the technical point of view, but also more realistic for the description, in mathematical terms, of the real-world system at hand. The reference mathematical structure is therefore Eq.~\eqref{eq:evol.disc}, which we rewrite adapting it to the present context:
\begin{equation}
    \frac{df_i}{dt}(t)=\sum_{k=1}^{n}\sum_{h=1}^{n}\eta_{hk}(t)\B_{hk}(i)f_h(t)f_k(t)
        -f_i(t)\sum_{k=1}^{n}\eta_{ik}(t)f_k(t), \quad i=1,\,\dots,\,n.
    \label{eq:evol.disc.onepop}
\end{equation}
Among the possible choices, we select a uniformly spaced wealth grid in the interval $D_u=[-1,\,1]$ with odd $n$:
\begin{equation}
    \begin{array}{c}
        I_u=\{u_1=-1,\,\dots,\,u_{\frac{n+1}{2}}=0,\,\dots,\,u_n=1\}, \\[3mm]
        u_i=\dfrac{2}{n-1}i-\dfrac{n+1}{n-1}, \quad i=1,\,\dots,\,n,
    \end{array}
    \label{eq:wealth.grid}
\end{equation}
agreeing that $u_i<0$ identifies a poor class whereas $u_i>0$ a wealthy one.

We next assume that the \textbf{encounter rate} $\eta_{hk}\geq 0$ is piecewise constant over the wealth classes:
\begin{equation}
    \eta_{hk}=
        \begin{cases}
            \eta_0 & \text{if\ } \abs{k-h}\leq\gamma \ \text{(competition)}, \\
            \mu\eta_0 & \text{if\ } \abs{k-h}>\gamma \ \text{(cooperation)},
        \end{cases}
    \label{eq:enc.rate}
\end{equation}
where $\eta_0>0$ is a constant to be hidden in the time scale and $0<\mu\leq 1$.

The \textbf{transition probabilities} $\B_{hk}(i)\in [0,\,1]$ are required to satisfy condition~\eqref{eq:sum.prob.discr}, which implies the conservation in time of the total number of active particles:
$$ N(t)=\sum_{i=1}^{n}f_i(t)=\text{constant}, \quad\forall\,t\geq 0, $$
plus an additional condition ensuring the conservation of the average wealth status of the population:
\begin{equation}
    \sum_{i=1}^{n}u_if_i(t)=\text{constant}, \quad\forall\,t\geq 0.
    \label{eq:const.wealth}
\end{equation}
This means that the interaction dynamics cause globally neither production nor loss of wealth, but simply its redistribution among the classes. We will denote by $U_0$ the average wealth status as fixed at the initial time:
$$ U_0:=\sum_{i=1}^{n}u_if_i(0). $$
By computing on Eq.~\eqref{eq:evol.disc.onepop}, it turns out that sufficient conditions for the fulfillment of \eqref{eq:const.wealth} are:
\begin{itemize}
\item symmetric encounter rate, i.e., $\eta_{hk}=\eta_{kh}$, $\forall\,h,\,k=1,\,\dots,\,n$;
\item transition probabilities such that
    \begin{equation}
        \sum_{i=1}^{n}u_i\B_{hk}(i)=u_h+\sigma_{hk}, \qquad \forall\,h,\,k=1,\,\dots,\,n,
        \label{eq:trans.prob.cons.U0}
    \end{equation}
where $\sigma_{hk}$ is an antisymmetric tensor, i.e., $\sigma_{hk}=-\sigma_{kh}$, $\forall\,h,\,k=1,\,\dots,\,n$.
\end{itemize}

Notice that the encounter rate \eqref{eq:enc.rate} is indeed symmetric. In order to explain condition~\eqref{eq:trans.prob.cons.U0}, let us consider preliminarily the particular case $\sigma_{hk}=0$ for all $h,\,k$. Then \eqref{eq:trans.prob.cons.U0} reduces to $\sum_{i=1}^{n}u_i\B_{hk}(i)=u_h$, which says that the expected wealth class of a candidate particle after an interaction coincides with its class before the interaction. Namely, interactions do not cause, in average, either enrichment or impoverishment, pretty much like a fair game. In the general case, Eq.~\eqref{eq:trans.prob.cons.U0} allows for fluctuations of the expected post-interaction wealth classes, over the pre-interaction one, however such that they globally balance: $\sum_{h,k=1}^{n}\sigma_{hk}=0$.

A possible set of transition probabilities describing cooperation/competition dynamics according to the distance between the interacting classes is, with minor modifications, that proposed in \cite{bertotti2008cla}:
\begin{eqnarray}
    && h=k
    	\begin{cases}
    		\B_{hh}(h)=1 \\
    		\B_{hh}(i)=0\ \forall\,i\ne h
    	\end{cases} \nonumber \\
    && h\ne k
    	\begin{cases}
            \begin{minipage}[c]{2.2cm}
                $\abs{k-h}\leq\gamma$ \\
                (competition)
            \end{minipage}
                \begin{cases}
                    h=1,\,n
                        \begin{cases}
                            \B_{hk}(h)=1 \\
                            \B_{hk}(i)=0\ \forall\,i\ne h
                        \end{cases} \\
                    h\ne 1,\,n
                        \begin{cases}
                            h<k
                                \begin{cases}
                                    k\ne n
                                        \begin{cases}
                                            \B_{hk}(h-1)=\alpha_{hk} \\
                                            \B_{hk}(h)=1-\alpha_{hk} \\
                                            \B_{hk}(i)=0\ \forall\,i\ne h-1,\,h
                                        \end{cases} \\
                                    k=n
                                        \begin{cases}
                                            \B_{hn}(h)=1 \\
                                            \B_{hn}(i)=0\ \forall\,i\ne h
                                        \end{cases}
                                \end{cases} \\
                            h>k
                                \begin{cases}
                                    k\ne 1
                                        \begin{cases}
                                            \B_{hk}(h)=1-\alpha_{hk} \\
                                            \B_{hk}(h+1)=\alpha_{hk} \\
                                            \B_{hk}(i)=0\ \forall\,i\ne h,\,h+1
                                        \end{cases} \\
                                    k=1
                                        \begin{cases}
                                            \B_{h1}(h)=1 \\
                                            \B_{h1}(i)=0\ \forall\,i\ne h
                                        \end{cases}
                                \end{cases}
                        \end{cases}
                \end{cases} \\[7mm]
            \begin{minipage}[c]{2.2cm}
                $\abs{k-h}>\gamma$ \\
                (cooperation)
            \end{minipage}
                \begin{cases}
                    h<k
                        \begin{cases}
                            \B_{hk}(h)=1-\alpha_{hk} \\
                            \B_{hk}(h+1)=\alpha_{hk} \\
                            \B_{hk}(i)=0\ \forall\,i\ne h,\,h+1
                        \end{cases} \\
                    h>k
                        \begin{cases}
                            \B_{hk}(h-1)=\alpha_{hk} \\
                            \B_{hk}(h)=1-\alpha_{hk} \\
                            \B_{hk}(i)=0\ \forall\,i\ne h-1,\,h,
                        \end{cases}
                \end{cases}
        \end{cases}
    \label{eq:table.games.1}
\end{eqnarray}
where it is assumed that interactions within the same class produce no effect.

The parameter $\alpha_{hk}\in[0,\,1]$ appearing in Eq.~\eqref{eq:table.games.1} has the following meaning:
\begin{itemize}
\item in case of competition, it is the probability that the candidate particle further increases or decreases its wealth if it is, respectively, richer or poorer than the field particle;
\item in case of cooperation, it is the probability that the candidate particle gains or transfers part of its wealth if it is, respectively, poorer or richer than the field particle.
\end{itemize}
This probability may be constant, like in the already cited work \cite{bertotti2008cla}, or, as we will assume throughout the remaining part of this paper, may depend on the wealth classes, e.g.,
\begin{equation}
	\alpha_{hk}=\frac{\abs{k-h}}{n-1}, 
	\label{eq:alpha}
\end{equation}
in such a way that the larger the distance between the interacting classes the more stressed the effect of cooperation or competition. Any proportionality constant can be transferred into a scaling of the time variable.

It can be checked, using Eq.~\eqref{eq:table.games.1}, that
$$\sum_{i=1}^{n}u_i\B_{hk}(i)=u_h+\epsilon_{hk}\alpha_{hk}\Delta{u}, $$
where $\Delta{u}=\frac{2}{n-1}$ is the constant step of the grid \eqref{eq:wealth.grid} while $\epsilon_{hk}$ may be either $-1$, or $0$, or $1$ (depending on $h,\,k$) and is antisymmetric. Since $\alpha_{hk}$ given by Eq.~\eqref{eq:alpha} is instead symmetric, the previous one turns out to be precisely condition~\eqref{eq:trans.prob.cons.U0} with $\sigma_{hk}=\epsilon_{hk}\alpha_{hk}\Delta{u}$, which guarantees that this model preserves the average wealth status of the system.

\begin{remark}
For wealth conservation purposes, the transition probabilities \eqref{eq:table.games.1} are such that the extreme classes never take part nor trigger social competition. Namely, a candidate particle in the class $h=1$ or $h=n$ can only stay in the same class after any interaction with whatever field particle. Correspondingly, a field particle in the class $k=1$ or $k=n$ can only cause a candidate particle to remain in its pre-interaction class, no matter what the latter is.
\end{remark}

The \textbf{critical distance} $\gamma$, taken constant in \cite{bertotti2008cla}, is here assumed to depend on the instantaneous distribution of the active particles over the wealth classes, in order to account for nonlinearly additive interactions. In more detail, the time evolution of $\gamma$ should translate the following phenomenology of (uncontrolled) social competition:
\begin{itemize}
\item in general, $\gamma$ grows with the number of poor active particles, thus causing larger and larger gaps of social competition. Few  wealthy active particles insist on maintaining, and possibly improving, their benefits;
\item in a population constituted almost exclusively by poor active particles $\gamma$ attains a value such that cooperation is inhibited, for individuals tend to be involved in a ``battle of the have-nots'';
\item conversely, in a population constituted almost exclusively by  wealthy active particles $\gamma$ attains a value such that competition is inhibited, because individuals tend preferentially to cooperate for preserving their common benefits.
\end{itemize}

Bearing these ideas in mind, we introduce the number of poor and wealthy active particles at time $t$:
$$ N^{-}(t)=\sum_{i=1}^{\frac{n-1}{2}}f_i(t), \qquad
    N^{+}(t)=\sum_{i=\frac{n+3}{2}}^{n}f_i(t). $$
Notice that, by excluding the middle class $u_\frac{n+1}{2}=0$ from both $N^{-}$ and $N^{+}$, we implicitly regard it as economically ``neutral''. Up to normalization over the total number of active particles, we have $0\leq N^\pm\leq 1$ with also $N^{-}+N^{+}\leq 1$, hence the quantity
$$ S:=N^{-}-N^{+}, $$
which provides a macroscopic measure of the \emph{social gap} in the population, is bounded between $-1$ and $1$. Given that, we now look for a quadratic polynomial dependence of $\gamma$ on $S$ taking into account the following conditions, which bring to a quantitative level the previous qualitative arguments:
\begin{itemize}
\item $S=S_0\Rightarrow\gamma=\gamma_0$, where $S_0$, $\gamma_0$ are a reference social gap and the corresponding reference critical distance, respectively;
\item $S=1\Rightarrow\gamma=n$, which implies that when the population is composed by poor particles only ($N^{-}=1$, $N^{+}=0$) the socio-economic dynamics are of full competition;
\item $S=-1\Rightarrow\gamma=0$, which implies that, conversely, when the population is composed by wealthy particles only ($N^{-}=0$, $N^{+}=1$) the socio-economic dynamics are of full cooperation.
\end{itemize}

\begin{figure}[!t]
\centering
\includegraphics[width=0.45\textwidth,clip]{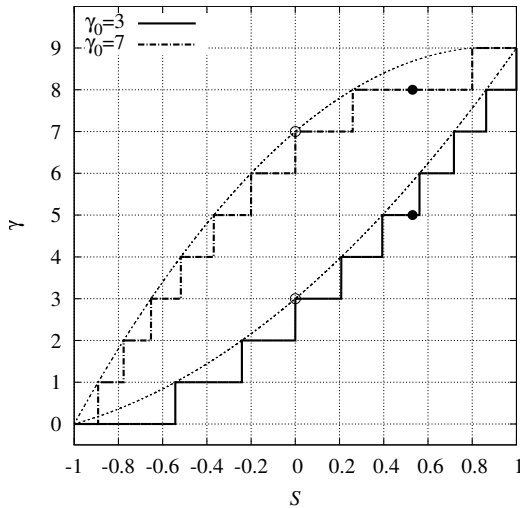}
\caption{The critical distance $\gamma$ vs. the macroscopic social gap $S$ as in Eq.~\eqref{eq:gamma_S0_0} for the two cases $\gamma_0=3,\,7$ and $n=9$ social classes. Empty bullets indicate the reference value $\gamma_0$ corresponding to the reference social gap $S_0=0$. Filled bullets indicate instead the actual initial critical distance $\gamma(t=0)$ corresponding to the actual initial social gap $S(t=0)$ for the case study addressed in Fig.~\ref{fig:blackswan1_U0-04}. Dotted lines, drawing the parabolic profile of function \eqref{eq:gamma_S0_0} without integer part, are plotted for visual reference.}
\label{fig:blackswan1_gamma_var}
\end{figure}

Considering further that only integer values of $\gamma$ are meaningful, for so are the distances between pairs of wealth classes, the resulting analytical expression of $\gamma$ turns out to be
\begin{equation}
    \gamma=\floor{\frac{2\gamma_0(S^2-1)-n(S_0+1)(S^2-S_0)}{2(S_0^2-1)}+\frac{n}{2}S},
    \label{eq:gamma}
\end{equation}
where $\floor{\cdot}$ denotes integer part (floor). In particular, if the reference social gap is taken to be $S_0=0$ (i.e., when $N^{-}=N^{+}$) then the expression of $\gamma$ specializes as (see Fig.~\ref{fig:blackswan1_gamma_var})
\begin{equation}
    \gamma=\floor{\frac{n-2\gamma_0}{2}S^2+\frac{n}{2}S+\gamma_0}.
    \label{eq:gamma_S0_0}
\end{equation}

\begin{remark}
Both $N^{-}$ and $N^{+}$ can be read, according to Eq.~\eqref{eq:moments.discr}, as zeroth-order weighted moments of the set of distribution functions $\{f_i\}_{i=1}^{n}$, with respective weights
$$  w(u)=
    \begin{cases}
        1 & \text{for\ } u\in [-1,\,0), \\
        0 & \text{for\ } u\in [0,\,1],
    \end{cases}
    \qquad \text{and} \qquad
    w(u)=
    \begin{cases}
        0 & \text{for\ } u\in [-1,\,0], \\
        1 & \text{for\ } u\in (0,\,1].
    \end{cases}
$$
Therefore, the dependence of $\gamma$ on $S$ introduces nonlinearly additive interactions in the transition probabilities $\B_{hk}(i)$.
\end{remark}

The evolution of the system predicted by the model depends essentially on the four parameters $n$ (the number of wealth classes), $\mu$ (the relative encounter rate for cooperation, cf. Eq.~\eqref{eq:enc.rate}), $U_0$ (the average wealth of the population), and $\gamma_0$ (the reference critical distance). The next simulations aim at exploring some aspects of the role that they play on the asymptotic configurations of the system. In more detail:
\begin{itemize}
\item $n=9$ and $\mu=0.3$ are selected;
\item two case studies for $U_0$ are addressed, namely $U_0=-0.4<0$ and $U_0=0$, in order to compare, respectively, the economic dynamics of a society in which poor classes dominate with those of a society in which the initial distribution of active particles encompasses uniformly poor and rich classes;
\item in addition, in each of the case studies above the asymptotic configurations for both constant and variable $\gamma$ are investigated, assuming, for duly comparison, that in the former the critical distance coincides with $\gamma_0$. Notice that a constant critical distance can be interpreted as an external control, for instance exerted by a Government, in order to supervise and regulate the wealth redistribution. The specific value of $\gamma$ can be related to more or less precautionary policies, depending on the allowed level of socio-economic competition. Particularly, $\gamma_0=3$, corresponding to a mainly cooperative attitude, and $\gamma_0=7$, corresponding instead to a strongly competitive attitude, are chosen.
\end{itemize}

\begin{figure}[!t]
\centering
\includegraphics[width=\textwidth,clip]{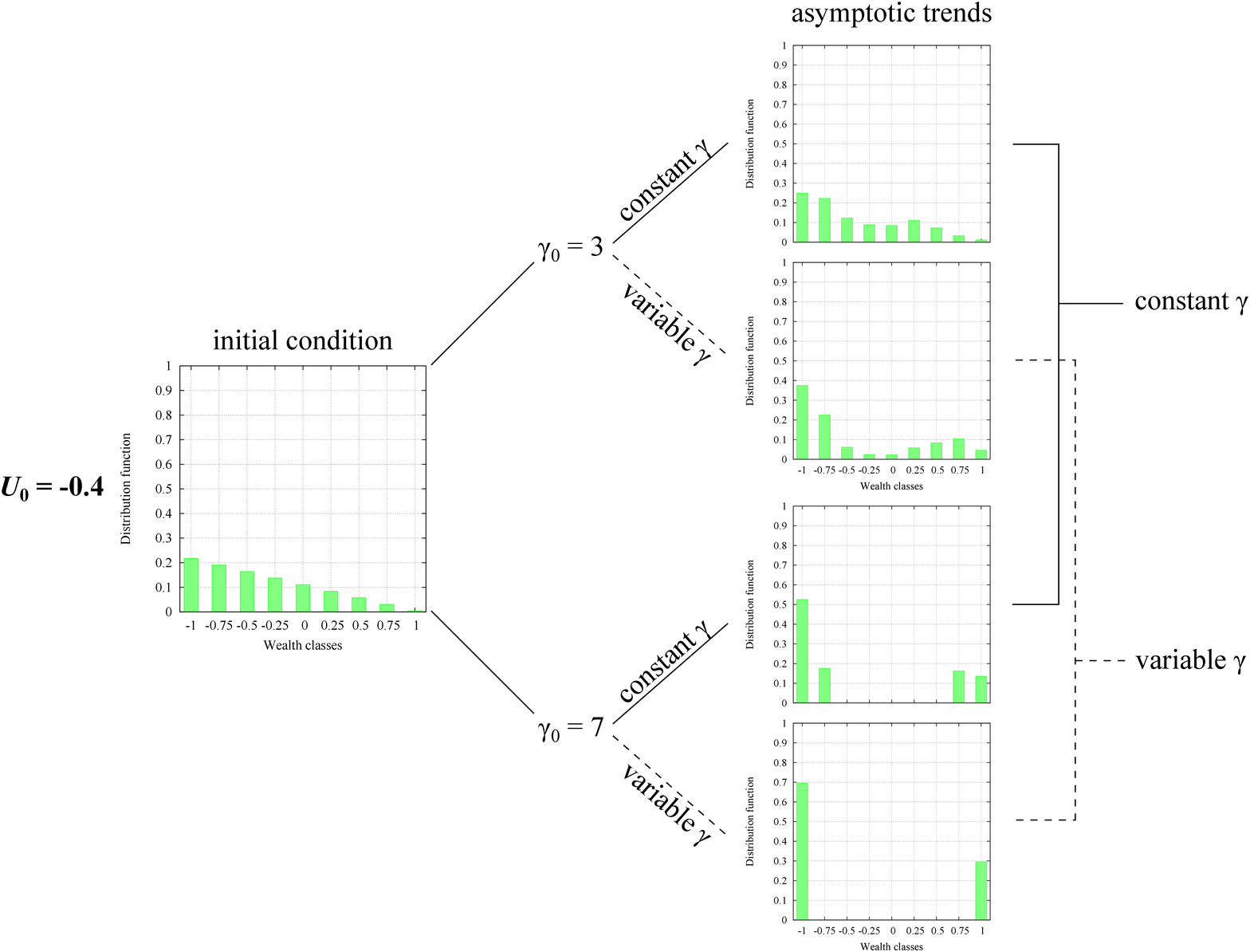}
\caption{Asymptotic distributions of active particles over wealth classes for $U_0=-0.4$.}
\label{fig:blackswan1_U0-04}
\end{figure}

Figure \ref{fig:blackswan1_U0-04} illustrates the asymptotic configurations in case of negative average wealth status $U_0$ (poor society). The model predicts, in general, a consolidation of the poorest classes. Nevertheless, in a basically cooperative framework ($\gamma_0=3$) a certain redistribution of part of the wealth is observed, which for controlled (viz. constant) $\gamma$ involves moderately poor and moderately rich classes whereas for uncontrolled  (viz. variable) $\gamma$ further stresses the difference between the poorest and the wealthiest classes. In fact, imposing a constant critical distance coinciding with the reference value $\gamma_0$ corresponds to forcing the society to behave as if the social gap were $S\equiv S_0=0$. On the other hand, the spontaneous attitude of the modeled society, in which the actual initial social gap computed from the given initial condition is $S(t=0)=\frac{8}{15}\approx 0.53>0$, is much more competitive than that implied by $\gamma_0=3$, as Fig.~\ref{fig:blackswan1_gamma_var} demonstrates. Analogous considerations can be repeated in a competitive framework ($\gamma_0=7$). Now the tendency is a strong concentration in the extreme classes, which in particular results in the consolidation of oligarchic rich classes which were nearly absent at the beginning.

\begin{figure}[!t]
\centering
\includegraphics[width=\textwidth,clip]{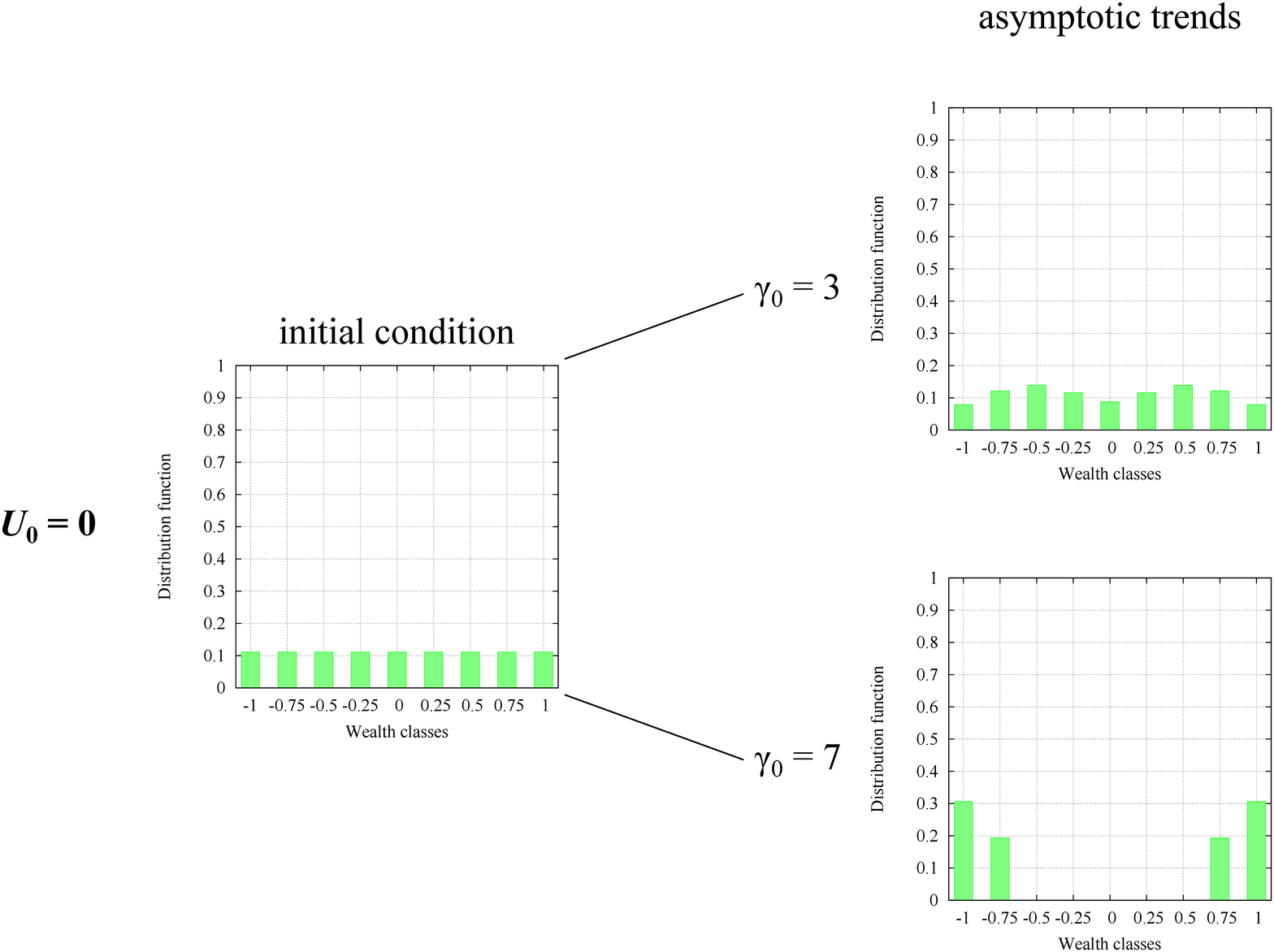}
\caption{Asymptotic distributions of active particles over wealth classes for $U_0=0$.}
\label{fig:blackswan1_U00}
\end{figure}

Figure \ref{fig:blackswan1_U00} illustrates instead the asymptotic trend in case of null average wealth status (economically ``neutral'' society). In this case there is no difference between the asymptotic configurations reached under constant and variable $\gamma$. Indeed, the initial symmetry of the distribution about the intermediate class $u_5=0$, which ensures $U_0=0$ and is preserved during the subsequent evolution, forces $S=0$, hence $\gamma=\gamma_0$, at all later times. In other words, controlled and spontaneous behaviors of the population coincide. The stationary configurations show a quite intuitive progressive clustering of the population in the extreme classes as the level of competition increases from $\gamma_0=3$ to $\gamma_0=7$.

As a general concluding remark, we notice that, even in a scenario of spontaneous/uncontrolled socio-economic dynamics (variable critical distance), none of the asymptotic configurations of the system seems to be properly identifiable as a Black Swan. On the other hand, the simple case studies addressed in this subsection are preliminary to the contents of the next subsection, which will focus on the joint effect of socio-economic and political dynamics. It is from the complex interplay between these two social aspects that Black Swans are mostly expected to arise.

\subsection{Modeling support/opposition to a Government}
\label{sect:support-opposition}
In this section we investigate how the welfare dynamics considered in Section~\ref{sect:welfare} can induce changes of personal opinions in terms of support/opposition to a certain political regime. In doing so, we will keep in mind recent results in the literature of Social Sciences; see for instance \cite{acemoglu2006eod,acemoglu2010tmd,acemoglu2011epi,acemoglu2010pcw,alesina2005wps}.

The mathematical structures to be used are those presented in Section~\ref{sect:transitions}, in particular the additional discrete microscopic variable $v$, which partitions the population into $m$ functional subsystems, represents the attitude of the individuals to the government. It is customary to use also for $v$ the uniformly spaced lattice
\begin{gather*}
    I_v=\{v_1=-1,\,\dots,\,v_{\frac{m+1}{2}}=0,\,\dots,\,v_m=1\}, \\[3mm]
    v_p=\dfrac{2}{m-1}p-\dfrac{m+1}{m-1}, \quad p=1,\,\dots,\,m,
\end{gather*}
agreeing that $v_1=-1$ corresponds to the strongest opposition whereas $v_m=1$ to the maximum support.

Mathematical models based on Eq.~\eqref{eq:evol.disc.2} are obtained by prescribing the encounter rate $\eta_{hk}^{pq}$ and the transition probabilities $\B_{hk}^{pq}(i,\,r)$. A very simple approach is proposed here, deferring to the next section a discussion on possible improvements.

For the \textbf{encounter rate} the same model given by Eq.~\eqref{eq:enc.rate} is assumed, according to the idea that encounters among active particle are mainly driven by the wealth state rather than by the difference of political opinion. Thus $\eta_{hk}^{pq}$ is independent of the functional subsystems that candidate and field particles belong to, $\eta_{hk}^{pq}=\eta_{hk}$. Notice that this amounts to disregarding political persuasion dynamics. The model could be made more precise, for instance, by allowing the encounter rate to depend on the proximity of political point of view of the interacting particles.

For the \textbf{transition probabilities} the following factorization on the output test state $(u_i,\,v_r)$ is proposed, relying simply on intuition:
$$ \B_{hk}^{pq}(i,\,r)=\bar{\B}_{hk}^{pq}(i)\hat{\B}_{hk}^{pq}(r), $$
where:
\begin{itemize}
\item $\bar{\B}_{hk}^{pq}(i)$ encodes the transitions of wealth class, which are further supposed to be independent of the political feelings of the interacting pairs: $\bar{\B}_{hk}^{pq}(i)=\bar{\B}_{hk}(i)$. For this term the structure given by Eq.~\eqref{eq:table.games.1} is used;
\item $\hat{\B}_{hk}^{pq}(r)$ encodes the changes of political opinion resulting from interactions. Coherently with the observation made above that political persuasion is neglected, so that political feelings originate in the individuals in consequence of their own wealth condition, this term is assumed to depend on the economic and political state of the candidate particle only: $\hat{\B}_{hk}^{pq}(r)=\hat{\B}_h^p(r)$.
\end{itemize}

In view of the special structure
$$ \B_{hk}^{pq}(i,\,r)=\bar{\B}_{hk}(i)\hat{\B}_h^p(r), $$
it turns out that sufficient conditions ensuring the conservation in time of both the total number of active particles and the average wealth status of the system are:
$$
\begin{cases}
    \dsum_{i=1}^{n}\bar{\B}_{hk}(i)=1, & \forall\,h,\,k=1,\,\dots,\,n \\[4mm]
    \dsum_{i=1}^{n}u_i\bar{\B}_{hk}(i)=u_h+\sigma_{hk},
    	& \forall\,h,\,k=1,\,\dots,\,n,\ \sigma_{hk}\ \text{antisymmetric} \\[4mm]
    \dsum_{r=1}^{m}\hat{\B}_h^p(r)=1, & \forall\,h=1,\,\dots,\,n,\ \forall\,p=1,\,\dots,\,m;
\end{cases}
$$
in particular, the first two statements are directly borrowed from Eq.~\eqref{eq:table.games.1}.

As far as the modeling of $\hat{\B}_h^p(r)$ is concerned, the following set of transition probabilities is proposed:
\begin{eqnarray}
	&& U_0<0,\ u_h<0
		\begin{cases}
			p=1
				\begin{cases}
					\hat{\B}_h^1(1)=1 \\
					\hat{\B}_h^1(r)=0\ \forall\,r\ne 1
				\end{cases} \\
			p>1
				\begin{cases}
					\hat{\B}_h^p(p-1)=2\beta \\
					\hat{\B}_h^p(p)=1-2\beta \\
					\hat{\B}_h^p(r)=0\ \forall\,r\ne p-1,\,p
				\end{cases}
		\end{cases} \nonumber \\
	&&
		\begin{minipage}[c]{23.5mm}
			\centering
			$U_0<0$, $u_h\geq 0$ \\
			or \\
			$U_0\geq 0$, $u_h<0$
		\end{minipage}
		\begin{cases}
			p=1
				\begin{cases}
					\hat{\B}_h^1(1)=1-\beta \\
					\hat{\B}_h^1(2)=\beta \\
					\hat{\B}_h^1(r)=0\ \forall\,r\ne 1,\,2
				\end{cases} \\
			1<p<m
				\begin{cases}
					\hat{\B}_h^p(p-1)=\beta \\
					\hat{\B}_h^p(p)=1-2\beta \\
					\hat{\B}_h^p(p+1)=\beta \\
					\hat{\B}_h^p(r)=0\ \forall\,r\ne p-1,\,p,\,p+1
				\end{cases} \\
			p=m
				\begin{cases}
					\hat{\B}_h^m(m-1)=\beta \\
					\hat{\B}_h^m(m)=1-\beta \\
					\hat{\B}_h^m(r)=0\ \forall\,r\ne m-1,\,m
				\end{cases}
		\end{cases}
		\label{eq:table.games.2} \\
	&& U_0\geq 0,\ u_h\geq 0
		\begin{cases}
			p<m
				\begin{cases}
					\hat{\B}_h^p(p)=1-2\beta \\
					\hat{\B}_h^p(p+1)=2\beta \\
					\hat{\B}_h^p(r)=0\ \forall\,r\ne p,\,p+1
				\end{cases} \\
			p=m
				\begin{cases}
					\hat{\B}_h^m(m)=1 \\
					\hat{\B}_h^m(r)=0\ \forall\,r\ne m,
				\end{cases}
		\end{cases} \nonumber
\end{eqnarray}
where $\beta\in[0,\,\frac{1}{2}]$ is a parameter expressing the basic probability of changing political opinion. According to Eq.~\eqref{eq:table.games.2}, transitions across functional subsystems are triggered jointly by the individual wealth status of the candidate particle and the average collective one of the population, in such a way that:
\begin{itemize}
\item poor individuals in a poor society ($U_0<0$, $u_h<0$) tend to distrust markedly the government policy, sticking
in the limit at the strongest opposition;
\item wealthy individuals in a poor society ($U_0<0$, $u_h\geq 0$) and poor individuals in a wealthy society ($U_0\geq 0$, $u_h<0$) exhibit, in general, the most random behavior. In fact, they may trust the government policy either because of their own wealthiness, regardless of the possibly poor general condition, or because of the collective affluence, in spite of their own poor economic status. On the other hand, they may also distrust the government policy either because of the poor general condition, in spite of their individual wealthiness, or because of their own poor economic status, regardless of the collective affluence;
\item wealthy individuals in a wealthy society ($U_0\geq 0$, $u_h\geq 0$) tend instead to trust earnestly the government policy, sticking in the limit at the maximum support.
\end{itemize}
In all cases, transitions are of at most one functional subsystem at a time, i.e., the output state of the candidate
particle is possibly in the higher or lower nearest subsystem.

In spite of a number of possible refinements of the model, some preliminary numerical simulations can be developed toward the main target of this paper. Specifically, we consider again the two cases corresponding to an economically neutral ($U_0=0$) and a poor ($U_0=-0.4<0$) society, assuming that the political feelings are initially uniformly distributed within the various wealth classes. The relevant parameters related to welfare dynamics are set as in Section~\ref{sect:welfare}. Additionally, the basic probability of changing political orientation is set to $\beta=0.4$, and $m=9$ functional subsystems are selected corresponding to as many levels of political support/opposition.

\begin{figure}[!t]
\centering
\includegraphics[width=\textwidth,clip]{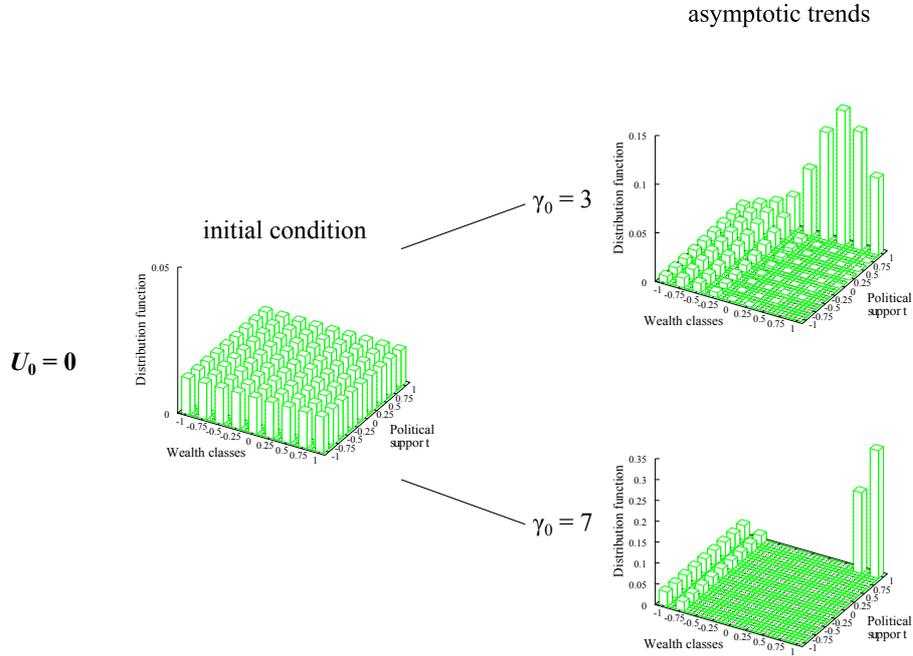}
\caption{Asymptotic distributions of active particles over wealth classes and political orientation for $U_0=0$.}
\label{fig:blackswan2_U00}
\end{figure}

\begin{figure}[!t]
\centering
\includegraphics[width=\textwidth,clip]{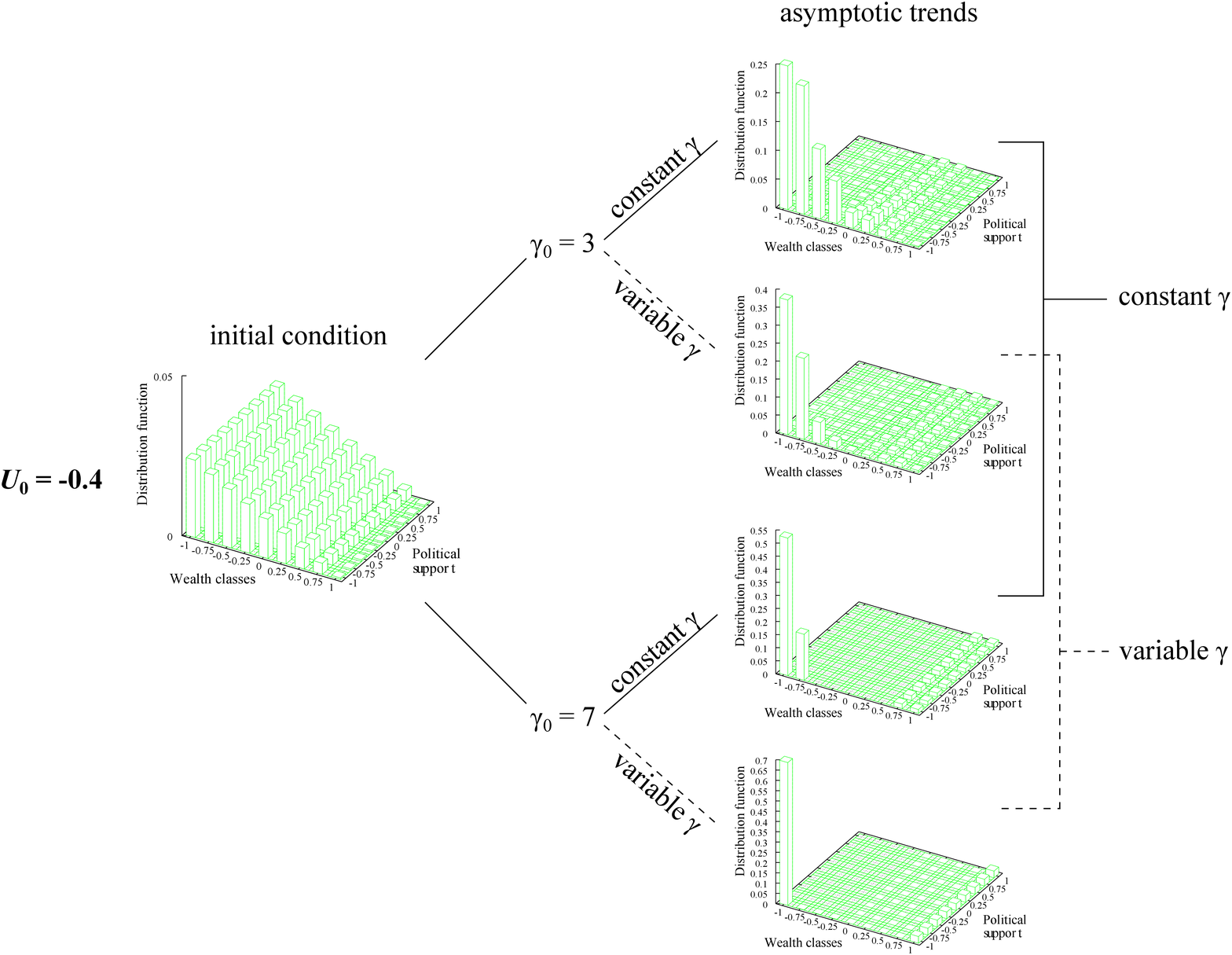}
\caption{Asymptotic distributions of active particles over wealth classes and political orientation for $U_0=-0.4$.}
\label{fig:blackswan2_U0-04}
\end{figure}

Figure~\ref{fig:blackswan2_U00}, referring to the case $U_0=0$, shows that in an economically  neutral society with uniform wealth distribution, such that controlled and uncontrolled welfare dynamics coincide, not only do wealthy classes stick at an earnest support to the Government policy, but also poor ones do not completely distrust them, especially in a context of prevalent cooperation among the classes ($\gamma_0=3$). Therefore, this example does not suggest the development of significant polarization in that society. On the other hand, Figure~\ref{fig:blackswan2_U0-04}, corresponding to the case $U_0=-0.4$, clearly shows a strong radicalization of the opposition. The model predicts indeed that, in such a poor society, poor classes stick asymptotically at the strongest opposition, whereas wealthy classes spread over the whole range of political orientations, however with a mild tendency toward opposition for the moderately rich ones (say, $u_5=0$, $u_6=0.25$, and $u_7=0.5$). The growth of political aversion is especially emphasized under uncontrolled welfare dynamics (i.e., variable critical distance $\gamma$), when the marked clustering of the population in the lowest wealth classes, due to a more competitive spontaneous attitude, entails in turn a clustering in the highest distrust of the regime.

\begin{remark}
The case studies discussed above indicate that an effective interpretation of the social phenomena under consideration requires a careful examination of the probability distribution over the microscopic states. Indeed, Figs.~\ref{fig:blackswan2_U00},~\ref{fig:blackswan2_U0-04} show entirely different scenarios, that might not be completely caught simply by average macroscopic quantities.
\end{remark}

\subsection{Looking for early signals of the Black Swan}
\label{sect:black.swan}
The simulations presented in the preceding sections have put in evidence that an unfair policy of welfare distribution can cause a radical surge of opposition to the regime. If this happens, intuitive consequences are, for example, strong social conflicts possibly degenerating into revolutions. Therefore, it is of some practical interest to look for early signals that may precede the occurrence of this situation.

To begin with, it is worth detailing a little more the expression \emph{Black Swan}, introduced in the specialized literature for indicating unpredictable events, which are far away from those generally observed by repeated empirical evidence. In \cite{taleb2007bsi}, a Black Swan is specifically characterized as follows:
\begin{quotation}
{\it ``A Black Swan is a highly improbable event with three principal  characteristics: It is unpredictable; it carries a massive impact; and, after the fact, we concoct an explanation that makes it appear less random, and more predictable, than it was.''}
\end{quotation}
and a critical analysis is developed about the failure of the existing  mathematical approaches to address such situations. In the author's opinion, this is due to the fact that mathematical models usually rely on what is already known, thus failing to predict what is instead unknown. It is worth observing that \cite{taleb2007bsi} is a rare example of research moving against the main stream of the traditional approaches, generally focused on well-predictable events. The book \cite{taleb2007bsi} had an important impact on the search of new research perspectives: for instance, it motivated applied mathematicians and other scholars to propose formal approaches to study the Black Swan, in an attempt to forecast conditions for its onset. In this context, the following remarks are in order.
\begin{itemize}
\item Mathematical models can serve either \emph{predictive} or \emph{exploratory} purposes. In the first case, they predict the evolution in time of the system for fixed initial conditions and parameters; in other words, they are used to simulate specific real-world situations of interest. In the second case, instead, they focus on the influence of initial conditions and free parameters on the overall evolution; namely, they are used to investigate the conditions under which desired or undesired behaviors may come up.
\item A successful modeling approach will eventually provide analytical methods for identifying the Black Swan, which in turn will be carefully defined in mathematical terms.
\item Individual behavioral rules and strategies are not, in most cases, constant in time due to the evolutionary characteristics of living complex system. Particularly, some parameters of the models, related to the interactions among the individuals, can change in time depending on the global state of the system. Such a variability may generate unpredictable events.
\item The qualitative analysis of social phenomena cannot be fully understood simply by average quantities. As already mentioned, the proper detail of mathematical description has to be retained over the microscopic states of the interacting subjects. Statistical distributions can serve such a purpose, while not forcing a one-by-one characterization of the agents.
\end{itemize}
It is plain that the mathematical search of the Black Swan can hardly rely on a purely macroscopic viewpoint. On the other hand, early signals of upcoming extreme events can be profitably sought at a macroscopic level, in order for them to be observable, hence recognizable, in practice.

Bearing in mind the previous remarks, we now provide some suggestions for the possible detection of a Black Swan within our current mathematical framework. The arguments that follow refer to closed systems in the absence of migrations, so that up to normalization the distribution functions can be regarded as probability densities.

Let us assume that a specific model, derived from the mathematical structures presented in Section~\ref{sect:compl.red}, has a trend to an asymptotic configuration described by stationary distribution functions $\{f_\asy^p\}_{p=1}^{m}$:
\begin{equation}
	\lim_{t\to +\infty}\norm{f_\asy^p-f^p(t,\,\cdot)}=0, \quad p=1,\,\dots,\,m,
	\label{eq:asympt}
\end{equation}
where $\norm{\cdot}$ is a suitable norm over the activity $u$, for instance
\begin{equation}
     \norm{g}_{1,w}=\intl_{D_u}\abs{g(u)}w(u)\,du \quad  \text{for\ } g\in L^1_{w}(D_u),
        \label{eq:norm.one}
\end{equation}
and $w:D_u\to\R_+$ is a weight function which takes into account the critical ranges of the activity variable. Equation~\eqref{eq:norm.one} is written for a continuous activity variable; its counterpart in the discrete setting is
$$ \norm{g}_{1,w}=\sum_{i=1}^{n}\abs{g(u_i)}w(u_i), $$
now valid for $g,\,w\in C^0(D_u)$. Alternative metrics can also be introduced depending on the phenomenology of the system at hand, which may need, for instance, either uniform or averaged ways of measuring the distance between different configurations.

In addition, let us assume that the modeled system is expected to exhibit a stationary trend described by some phenomenologically guessed distribution functions $\{\tilde{f}_\asy^p\}_{p=1}^{m}$. In principle, such expected distribution have to be determined heuristically for each specific case study, as we will see in the following.

Inspired by Eq.~\eqref{eq:asympt}, we define the following time-evolving distance $\dbs$ (the subscript ``BS''  standing for Black Swan):
\begin{equation}
    \dbs(t):=\max_{p=1,\,\dots,\,m}\norm{\tilde{f}_\asy^p-f^p(t,\,\cdot)},
    \label{eq:dbs}
\end{equation}
which, however, will generally not approach zero as time goes by for the heuristic asymptotic distribution does not translate the actual trend of the system. Using the terminology introduced in \cite{scheffer2009ews}, this function can be possibly regarded as one of the \emph{early-warning signals} for the emergence of critical transitions to rare events, because it may highlight the onset of strong deviations from expectations.

\begin{remark}
Specific applications may suggest other distances different from \eqref{eq:dbs}. For instance, linear or quadratic moments might be taken into account. We consider that extreme events are likely to be generated by the interplay of different types of dynamics, a fact that should be reflected in the choice of appropriate metrics.
\end{remark}

\begin{figure}[!t]
\centering
\includegraphics[width=0.45\textwidth,clip]{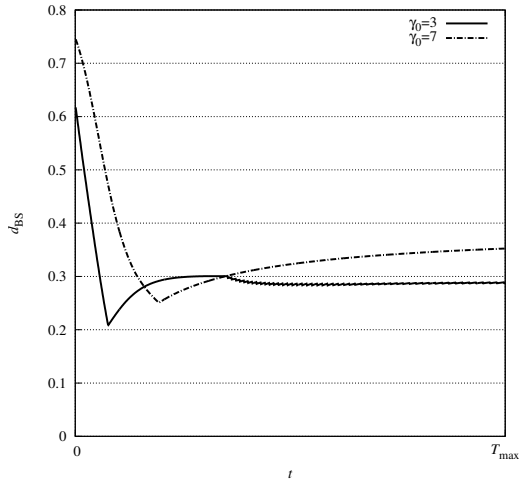}
\caption{The mapping $t\mapsto\dbs(t)$ computed in the case studies with variable $\gamma$ illustrated in Fig.~\ref{fig:blackswan2_U0-04}, taking as phenomenological guess the corresponding asymptotic distributions obtained with constant $\gamma$.}
\label{fig:dBS}
\end{figure}

It is interesting to examine the time evolution of the distance $\dbs$ with reference to the case study $U_0=-0.4$ with variable critical distance addressed in Section~\ref{sect:support-opposition}. A meaningful choice of the expected asymptotic distribution is, for both $\gamma_0=3$ and $\gamma_0=7$, the one resulting from the corresponding dynamics with constant critical distance. Reference is to a situation in which a government underestimates the role played by free interaction rules, presuming that the actual dynamics do not differ substantially from those observed under imposed rules. Figure~\ref{fig:dBS} shows the qualitative trends of the mapping $t\mapsto\dbs(t)$: an initial decrease of the distance, which may suggest a convergence to the guessed distribution, hence apparently a confirmation of the government's conjecture, is then followed by a sudden increase (notice the singular point in the graph of $\dbs$) toward a nonzero steady value, which ultimately indicates a deviation from the expected outcome. Such a turnround is possibly a macroscopic signal that a Black Swan is about to appear. However, in order to get a complete picture, the average gross information delivered by $\dbs(t)$ has to be supplemented by the detailed knowledge of the probability distribution over the microscopic states, which is the only one able to properly distinguish between lower ($\gamma_0=3$) and higher ($\gamma_0=7$) radicalization of political feelings when welfare dynamics are left to individual selfishness \cite{acemoglu2006eod}.

\section{Critical analysis}
\label{sect:discussion}
In this paper we have considered the problem of modeling complex systems of interacting individuals, focusing in particular on the ability of the models to predict the onset of rare events that cannot be generally foreseen on the basis of past empirical evidences. The results presented in the preceding sections are encouraging, yet a critical analysis is necessary in order to understand how far we still are from the challenging goal of devising suitable mathematical tools for studying the emergence of highly improbable events. We feel confident that a first step is that direction has been made in this paper. On the other hand, we do not naively pretend that the ultimate target has been met. With the aim of contributing to further improvements, we propose in the following some considerations about specific problems selected according to our scientific bias. Hopefully, this selection addresses key issues of the theory.

\paragraph*{Mathematical tools for complex systems} The leading idea of the present paper is that the modeling approach to socio-economic and political systems, where individual behaviors can play a relevant role on the collective dynamics, needs to consider the latter as living complex systems. This implies characterizing them in terms of qualitative complexity issues proper of Social Sciences (cf. Section~\ref{sect:compl.asp}), that have then to be translated in the mathematical language. The mathematical tools presented in the preceding sections are potentially able to capture such issues, taking advantage of a procedure of complexity reduction in modeling heterogeneous behaviors and expression of strategies. Yet, no matter how promising this approach may appear, we are not pretending that it is sufficient ``as-is'' for chasing the Black Swan. Instead, models such as those reported in Section~\ref{sect:case.studies} can provide a detailed analysis of events whose broad dynamics are rather well understood. Furthermore, simulations contribute to put in evidence the role of some key parameters and can indicate how to devise external actions in order to eventually obtain a specific behavior of the society under consideration.

\paragraph*{Modeling interplays toward the Black Swan} Based on the preliminary results that we have obtained, we believe that rare events can only be generated by several concomitant causes. In this paper we have addressed the interplay between welfare dynamics and the level of consent/dissent to the policies of a government, with the aim of detecting the onset of the opposition to a certain regime. Numerical simulations have indicated that a strong opposition can result from specific conditions, such as a poor average wealth status administered under a welfare policy leaving freely to the market the rules of cooperation and competition, without any action by the central government. On the other hand, the social dissent is attenuated if the government has some control on the welfare dynamics, for instance if it is able to keep an acceptable level of cooperation within the population in spite of the spontaneous competitive behavior induced by the poor collective condition. Of course, the investigation can be further refined by taking into account additional causes. According to the methodological approach proposed in this paper, the latter imply partitioning the population in additional functional subsystems.

\paragraph*{A naive interpretation of the recent events in North Africa} The contents and findings of the present work inspire some considerations, no matter how naive they may appear, about the recent conflicts in North Africa countries. First of all, we notice that the latter feature all issues which, according to Taleb's definition \cite{taleb2007bsi}, characterize a Black Swan. Such events were indeed not expected, but their social impact has been definitely important. Furthermore, once they happened it actually seemed that they could have been foreseen, for instance it was argued that a wiser political management of welfare would have limited, or even avoided, their occurrence. Coming to the conceivable predictive ability of the mathematical approach presented in this paper, we remark that the dynamics just recalled are indeed accounted for by the models proposed in Section~\ref{sect:case.studies}. Actually, we are well aware that the events we are reasoning upon were generated by numerous concomitant causes other than simply welfare dynamics. Therefore, while not claiming to have exhaustively tackled them, we hope to have provided a significant contribution to the problem of detecting early signals for such critical events.

\paragraph*{Further generalizations of the model} The modeling approach can be generalized for instance by considering the case of open systems, in which external actions can significantly modify both individual and collective system dynamics. Analytical properties of formal mathematical structures, which may be profitably employed to address such issues, have been studied in \cite{arlotti2012cid}, however models specifically targeted at real-world problems are not yet available. Other interesting applications concern the case of several interacting societies, for example the study of how and when a domino effect, like the one recently observed in the aforesaid North Africa countries, can arise. Even more challenging appears to be the generalization of the model to large social networks \cite{barabasi1999mft,vegaredondo2007csn}. Recent studies, among others \cite{bastolla2009amn,rand2011dsn}, indicate that the role and structure of the networks can act as additional inputs for determining the predominance of either cooperation or competition. However, exploring this issue requires a substantial development of the mathematical structures presented in this paper.

\paragraph*{Analytical problems} The qualitative analysis of models of the kind presented in this paper generates interesting analytical problems. As a matter of fact, showing the existence and uniqueness of solutions to the initial value problem is not a difficult task, because one can exploit the conservation of the total number of individuals and of their average wealth status. For linearly additive interactions the proof can be obtained by a simple application of fixed point theorems in a suitable Banach space, see \cite{arlotti1996snc}; the generalization to nonlinearly additive interactions has been recently proposed in the already cited paper \cite{arlotti2012cid}. Far more challenging is the analysis, for arbitrary numbers of wealth classes and functional subsystems, of existence and stability of asymptotic configurations, which are at the core of the practical implications of the model. Simulations suggest that, for a given initial condition, the system reaches a unique asymptotic configuration in quite a broad range of parameters, but the existing literature still lacks precise analytical results able to confirm or reject such a conjecture. Some preliminary insights, however confined to linearly additive interactions, can be found in \cite{arlotti1996snc}, that hopefully may serve as a starting point for more general proofs.

\bibliographystyle{plain}
\bibliography{BnHmTa-blackswan}

\end{document}